\documentclass[letterpaper,aps,prl,twocolumn,showpacs]{revtex4}
\usepackage{graphicx,bm}
\usepackage{amsmath,amssymb}

\begin{document}

\title{Andreev bound states in iron pnictide superconductors}
\author{Wen-Min Huang$^1$}
\author{Hsiu-Hau Lin$^{1,2}$}
\affiliation{$^1$Department of Physics, National Tsing Hua University, Hsinchu 30013, Taiwan}
\affiliation{$^2$Physics Division, National Center for Theoretical Sciences, Hsinchu 30013, Taiwan}

\date{\today}
%74.78.-w	Superconducting films and low-dimensional structures
%74.20.-z  Theories and models of superconducting state
%74.45.+c	Proximity effects; Andreev effect; SN and SNS junctions

\begin{abstract}
Recently, Andreev bound states in iron pnictide have been proposed as an experimental probe to detect the relative minus sign in the $s_\pm$-wave pairing. While previous theoretical investigations demonstrated the feasibility of the approach, the local density of states in the midgap regime is small, making the detection hard in experiments. We revisit this important problem from the Bogoliubov-de Gennes Hamiltonian on the square lattice with appropriate boundary conditions. Significant spectral weights in the midgap regime are spotted, leading to easy detection of the Andreev bound states in experiments. Peaks in the momentum-resolved local density of states appear and lead to enhanced quasiparticle interferences at specific momenta. We analyze the locations of these magic spots and propose they can be verified in experiments by the Fourier-transformed scanning tunneling spectroscopy.
\end{abstract}

\maketitle

Discovery of superconductivity in LaFeAsO$_{1-x}$F$_x$\cite{Kamihara08} and related iron pnictides\cite{Rotter08,Tapp08,Hsu08} recently has reignited intense investigations on the unconventional superconductivity\cite{Norman08,Hosono09}. As a first step toward the understanding of the underlying pairing mechanism, it is important to know the pairing symmetry in iron pnictides. Recent experiments on the nuclear magnetic resonance\cite{Grafe08, Nakai08}, the penetration depth\cite{ Matano08,Luetkens08,Fletcher09,Hicks09} and the specific heat\cite{Gordon09} reveal the signature of the nodes in the gap. However, some other experimental measurements\cite{Terasaki09,Hashimoto09,Malone09,Luo09} show the opposite. In particular, the angular-resolved photoemission spectroscopy\cite{Zhao08, Ding08, Kondo08,Terashima09} clearly shows the presence of the gap at all points on the Fermi surface. Thus, the pairing symmetry in iron pnictides remains controversial at the point of writing.

Band structure calculations\cite{Singh08, Kuroki08} show that the low-energy excitations mainly come from the two dimensional FeAs layers with Fermi surface composed of two disconnected hole and electron pockets. Nesting between the hole and electron pockets causes the magnetic ordering before the superconductivity sets in, hinting that the spin-fluctuation induced pairing\cite{Terasaki09,Terashima09,Mazin08} is plausible in these materials. Furthermore, renormalization group approaches\cite{Eremin08,FWang09,Honerkamp09} show that the leading contender in various pairing symmetries is the extended $s$-wave (or $s_{\pm}$-wave) pairing state where the Fermi surface is fully gapped but the gap functions for the electron and hole pockets show opposite signs. The unconventional sign flip in the gap functions initiates intense studies\cite{Wang09,Onari09,Golubov09,Zhang09,Ghaemi09} to reveal its role in the superconducting phase and also potential connection to the magnetic instabilities.

Andreev bound states (ABS)\cite{Ghaemi09} serves as an indirect probe for the relative minus sign in the $s_{\pm}$-wave paring. In contrast to the $d$-wave pairing symmetry in cuprates\cite{Hu94}, ABS in iron pnictides appears at finite energy with dispersion depending on the microscopic details. A significant drawback as an experimental probe is the lack for significant peak in the local density of states\cite{Ghaemi09} in the midgap regime. In this Letter, we revisit this problem but treating the lattice effects in iron pnictides carefully with appropriate boundary conditions. Note that the work function near the sample boundary gives rise to an effective potential and shall be included. In contrast to previous results, we find significant peaks in the momentum-resolved local density of states and large spectral weights in the midgap regime. In addition, we also predict the existence of the magic spots due to quasiparticle interference that can be verified in experiments by the Fourier-transformed scanning tunneling spectroscopy (FT-STS)\cite{Hoffman02,McElroy03}.     

\begin{figure}[t!]
\centering
\includegraphics[width=7.2cm]{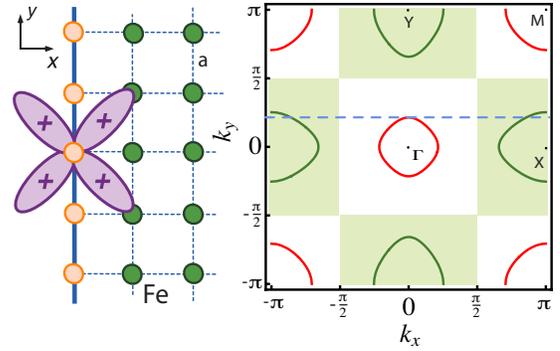}
\caption{(Color online) A semi-infinite square lattice for iron pnictide with a flat edge. The surface potential $V_0$ locates along the boundary sites colored in orange. The Fermi surface of the two-band model with tight-binding parameters $t_1=-1$, $t_2=1.3$, $t_3=t_4=0.85$ and $\mu=1.55$ is shown. The pairing potential on the lattice and in the momentum space is shown here and the shaded area in the Brillouin zone denotes the sign change in the gap function. The blue dashed line stands for the transverse momentum where the hole pocket near the $\Gamma$ point vanishes. }\label{FS}
\end{figure}

To elaborate on the details, we start with the Bogoliubov-de Gennes (BdG) Hamiltonian for the two-orbital model with the $s_{\pm}$-wave pairing on the square lattice, 
\begin{eqnarray}\label{BdG}
&&\hspace{-0.4cm}H_{BdG}=\sum_{a,b}\sum_{ \bm{r}, \bm{r}';\sigma}\left[t_{ab}(\bm{r}, \bm{r}')-\mu\hspace{0.1cm}\delta_{a,b}\delta_{\bm{r}, \bm{r}'}\right]c^{\dag}_{a\sigma}( \bm{r})c_{b\sigma}( \bm{r}')\nonumber
\\ &&\hspace{1.1cm}+\sum_{a,b}\sum_{\bm{r},\bm{r}'}\left[\Delta_{ab}^*( \bm{r}, \bm{r}')c_{a\uparrow}(\bm{r})c_{b\downarrow}(\bm{r}')+{\rm h.c.}\right],
\end{eqnarray}
where $\sigma$ labels the spins and $a,b=X,Y$ denote the $d_{XZ}$ and $d_{YZ}$ orbitals respectively. The non-vanishing hopping amplitudes $t_{ab}$ on the square lattice include the intraorbital hopping $t_{XX}(x\pm 1,y)=t_{YY}(x,y\pm 1)=-t_1$, $t_{XX}(x,y\pm 1)=t_{YY}(x\pm 1,y)=-t_2$, $t_{XX/YY}(x\pm 1,y\pm 1)=t_{XX/YY}(x\mp 1,y\pm 1)=-t_3$ and the interorbital hopping $t_{XY}(x\pm 1,y\pm 1)=-t_{XY}(x\mp 1,y\pm 1)=t_4$. The lattice constant is set to unity $a=1$ through out the paper. With appropriate parameters, it was shown that the two-band model capture the essential features of the realistic band structure\cite{Zhang081, Lee091}. The Fermi surface contains two electron pockets at the $X$ and $Y$ points and two hole pockets at the $\Gamma$ and $M$ points in the Brillouin zone as shown in Fig.~\ref{FS}. The gap function with the $s_{\pm}$-wave symmetry is described by the intraorbital pairing $\Delta_{XX/YY}(x\pm 1,y\pm 1)=\Delta_{XX/YY}(x\mp 1,y\pm 1)=\Delta_0$ on the square lattice shown in the left panel of the Fig.~\ref{FS}. 

The translational invariance along the $y$ direction makes the corresponding momentum $k_y$ a good quantum number. The open boundary ruin the translation invariance in the $x$ direction but the solution can be constructed by applying the generalized Bloch state\cite{Lin05,Huang08}, $\Phi_z(x)=\Phi z^x$, where $z$ is the complex eigenvalue of the displacement operator, $D \Phi_z(x)= z \Phi_z(x)$. The BdG Hamiltonian can be decomposed into different $k_y$ momentum sectors and the eigenvalue problem is greatly simplified to $(H_z-\epsilon{\bf{1}})\Phi=0$, or in explicit matrix form
\begin{eqnarray}\label{Harper}
\left(\begin{array}{cccc}T_{11}-\epsilon &T_{12} & \Delta_{11} & 0\\ T_{21} & T_{22}-\epsilon & 0 & \Delta_{22} \\ \Delta_{11} & 0 & -T_{11}-\epsilon & -T_{12} \\0 & \Delta_{22} & -T_{12} & -T_{22}-\epsilon\end{array}\right)\hspace{-0.2cm}\left(\begin{array}{c} \phi_{1\uparrow} \\\phi_{2\uparrow} \\\phi_{1\downarrow} \\\phi_{2\downarrow} \end{array}\right)\hspace{-0.1cm}=0,
\end{eqnarray}
where $T_{11}=t_{11}/z-\left(\mu_1+\mu\right)+t_{11}z$, $T_{22}=t_{22}/z-\left(\mu_2+\mu\right)+t_{22}z$, $T_{12}=-it_{12}/z+it_{12}z$ and $\Delta_{11}=\Delta_{22}=\Delta_1/z+\Delta_1z$. The effective hopping amplitudes and the gap functions are $t_{11}=-t_1-2t_3\cos k_y$, $\mu_1=-2t_2\cos k_y$, $t_{22}=-t_2-2t_3\cos k_y$, $\mu_2=-2t_1\cos k_y$, $t_{12}=2t_4\sin k_y$, and $\Delta_1=2\Delta_0\cos k_y$. For given $k_y$ and $\epsilon$, one can solve for eight solutions of $z$ with corresponding eigenvectors $\Phi$. Unit modulus $|z|=1$ solutions are the usual plane waves and $|z|<1$ corresponds to the evanescent modes due to the presence of the open boundary. Note that, if $z$ is a solution, we can show that $1/z$ is also a solution. Thus, we expect only four $|z|<1$ evanescent modes.

The general solution for the evanescent modes is $\Psi(x)=\sum_{\gamma=1}^4a_{\gamma}\Phi_{\gamma}(x)$ and should satisfies the boundary conditions. Due to the work function near the sample edge, we introduce an additional $V_0$ to incorporate the rise of the potential near the open boundary. After some algebra, boundary conditions can be casted into the matrix form,
\begin{eqnarray}\label{BC1}
\bm{B}_c\Psi(-1)+\bm{V}\Psi(0)=0,
\end{eqnarray}
with 
\begin{eqnarray}
\bm{B}_c\hspace{-0.05cm}=\hspace{-0.1cm}\left(\hspace{-0.2cm}\begin{array}{cccc}t_{11} & -it_{12} & \Delta_{1} & 0\\ -it_{12}& t_{22} & 0 & \Delta_{1} \\ \Delta_{1} & 0 & -t_{11} & it_{12} \\0 & \Delta_{1} & it_{12} & -t_{22}\end{array}\hspace{-0.1cm}\right),\bm{V}\hspace{-0.05cm}=V_0\left(\begin{array}{cc}\bf{1} & \bf{0} \\ \bf{0} & -\bf{1}\end{array}\hspace{-0.1cm}\right)\hspace{-0.05cm},
\end{eqnarray}
where $\bf{1}$ is a $2\times2$ identity matrix. Substituting the wave function $\Psi(x)$ into Eq.~(\ref{BC1}), the boundary conditions reduce to the homogeneous equation, $\left(\bm{B}_c\bm{N}'+\bm{V}\textbf{N}\right)A=0$, where $A=\left(a_1,a_2,a_3,a_4\right)$, $\bm{N}'=\left(\frac{1}{z_1}\Phi_1,\frac{1}{z_2}\Phi_2,\frac{1}{z_3}\Phi_3,\frac{1}{z_4}\Phi_4\right)$ and $\bm{N}=\left(\Phi_1,\Phi_2,\Phi_3,\Phi_4\right)$. If this homogeneous equation hosts nonzero solutions, the determinant must equal zero, $\det\left[\bm{B}_c\bm{N}'+\bm{V}\bm{N}\right]$=0.

\begin{figure}[t!]
\centering
\includegraphics[width=7.6cm]{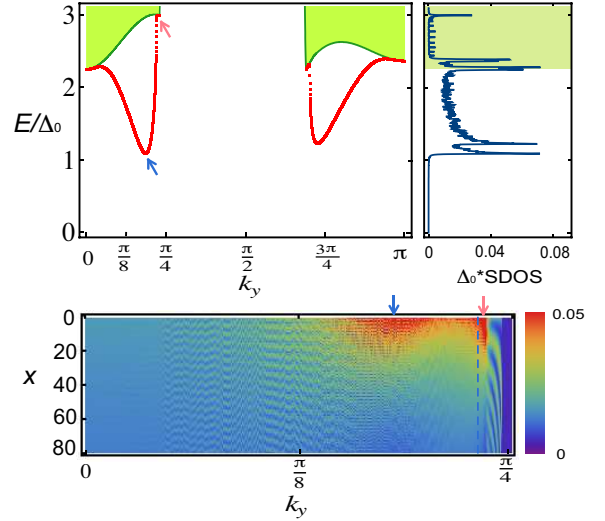}
\caption{(Color online) Dispersion of ABS with $\Delta_0=0.004$ (corresponding to the realistic values of $\sim15 meV$ for the superconducting gap) is shown in the upper-left panel. The surface potential $V_0/|t_1|=0.16$ is chosen here. The green shaded areas denote the continuum from the bulk states. The surface density of states is plotted in the upper-right panel and the momentum-solved probability density is shown in the lower panel. The blue dashed line stands for the transverse momentum where the hole pocket vanishes as in Fig.~\ref{FS}.}\label{LDOS}
\end{figure}

To obtain the full dispersion of the ABS, we need to solve for the generalized Bloch state for each momentum $k_y$. Since the dispersion is symmetric around $k_y=0$, we only show the portion between $k_y=0$ and $k_y =\pi$ in Fig.~\ref{LDOS}. The green shaded area represents the quasiparticle states in the bulk. The continuum is separated into two sectors: the left part (with $0 \leq k_y \lesssim \pi/4$) corresponds to the Fermi surface near the $\Gamma$ and $X$ points in the Brillouin zone while the right part for the regime close to the $Y$ and $M$ points. In the midgap regime below the continuum, the dispersive ABS is plotted as the red curve. Since we do not linearize the energy spectrum near the Fermi surface, our approach naturally eliminate any artifacts in the vicinity where the ABS merges into the continuum. This is clearly demonstrated that the dispersion glues back to the bottom of the continuum without any singularity near $k_y=0, \pi$. The most important feature of the dispersion is the dips located at the transverse momenta $q_1 \simeq \pm 0.19 \pi$ (near the $\Gamma/X$ points) and $q_2 \simeq \pm 0.72\pi$ (near the $Y/M$ points). These dip structures give rise to large spectral weights inside the midgap regime and also produce robust magic spots in quasiparticle interferences.

First of all, the local density of states near the boundary, i.e. surface density of states (SDOS), can be computed numerically $D_s(E)=\frac{1}{L_y}\sum_{k_y}|\Psi(x=0,k_y)|^2\delta[E-\epsilon(k_y)]$. The results exhibits several peaks as shown in the right panel of Fig.~\ref{LDOS}. However, in realistic situations, impurity scattering is unavoidable and those peaks living inside the continuum (green shaded area) hybridize with the plane-wave states. Thus, we only focus on the midgap regime where the evanescent modes are protected. Clearly, there is significant spectral weight transfer into the midgap regime in sharp contrast to the previous studies. Two significant peaks, arisen from the dips in the dispersion, can be detected in tunneling junctions or related experimental setups. Compared with previous approaches, two ingredients seem to explain the major differences: the band structure is kept without linearization (beyond the Andreev equations) and the jump of the work function near the open boundary is included. With the large spectral weight transfer and also the pronounce peaks inside the midgap regime, the ABS can be detected experimentally and serves as an indirect probe for the pairing symmetry.

The momentum resolved probability density $|\Psi(x,k_y)|^2$ near the $\Gamma/X$ points can also be computed numerically as shown in the bottom part of Fig.~\ref{LDOS}. For clarity, we only show the momentum range $0 \leq k_y \leq \pi/4$ near the $\Gamma/X$ points. In the vicinity of $k_y=\pi/4$, there is no ABS and the probability density vanishes. A general trend is manifest: the ABS is more localized near the boundary when the dispersion dives away from the continuum into the midgap regime. Let us concentrate on the peaks near the open boundary (red areas near $x=0$) indicated by the blue and pink arrows first. When comparing with the dispersion, it is clear that the most dominant feature in the probability density profile occurs at the dip of the dispersion (indicated by the blue arrow). In fact, by integrating over all momenta, it is easy to check that the peaks in SDOS come from the dips in the dispersion. Another peak (indicated by the pink arrow) in the probability density profile arises due to the Van Hove singularity. Note that the blue dashed line corresponds to the transverse momentum where the hole packet near the $\Gamma$ point disappears. Meanwhile, the electron pocket near the $X$ point is almost empty as well. Thus, the density of states to form ABS is greatly enhanced and leads to the peak structure in the probability density. However, since the energy lies inside the overall continuum, it is unlikely to remain sharp.

In addition to the peak features, it is also interesting to explore the quantum interference patterns in momentum resolved probability density. Close to the blue dashed line, stripe oscillations appear and extend into the bulk regime. The enhanced interference pattern comes from the fact that the solutions for $z_i$ are almost real and thus make the oscillation length longer and visible. On the other hand, near $k_y=0$ regime, the solutions for $z_i$ are generically complex. Linear combinations of the four solutions cancel out the quantum interferences and make the probability density profile more or less smooth.

In FT-STS, one can scan the midgap regime and looks for magic spots in the Fourier-transformed spectroscopy data. Gradually increasing the energy, we will hit the first peak in SDOS as shown in Fig.~\ref{LDOS}. The dominant quasiparticle scattering comes from $q_1$ and leads to an enhanced interference peak at the transfer momentum $Q_1 = \pm 2 q_1 \simeq \pm 0.38 \pi$. As we further increase the energy, the second peak in the LDOS reigns. The quasiparticle scattering should lead to another magic spot at the transfer momentum $Q_2 = \pm 2 q_2 \simeq \pm 1.44 \pi$. However, since there are also ABS at other momenta, we expect the magic spots at $Q_2$ is likely to be weaker than those at $Q_1$. 

\begin{figure}[t!]
\centering
\includegraphics[width=7.2cm]{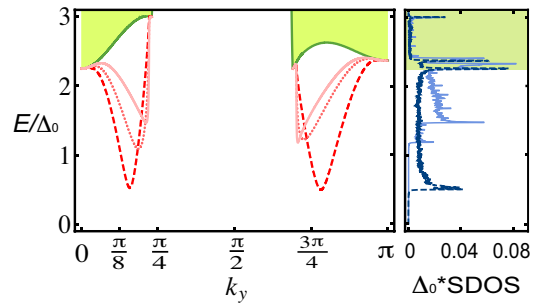}
\caption{(Color online) The dispersions of ABS with different surface potentials $V_0/|t_1|=0.08$, $0.16$, $0.40$, are showed as solid, dotted, dashed lines respectively. On the right-hand side, the surface density of states with $V_0/|t_1|=0.08$ and $0.40$ are plotted as light solid and dark dashed lines for comparison.
 }\label{differentV}
\end{figure}

Since the ABS occurs at finite energy, it depends on microscopic details. Therefore, it is important to explore how the above predictions change when the boundary potential $V_0$ varies. We show the ABS dispersions with several reasonable choices for the boundary potential $V_0$ in Fig.~\ref{differentV}. Clearly, the detail shape of the dispersion is sensitive to the boundary potential. As $V_0$ increases, the dispersion dives further into the midgap regime. However, several generic features remain robust : the spectral weight transfer inside the midgap regime is significant and there exists a sharp peak in SDOS. Therefore, the predictions in previous paragraphs remain qualitatively correct but may suffer some quantitative corrections. 

In summary, we develop a lattice-model approach for Andreev bound states in iron pnictides and find it can be detected in experiments as an indirect probe for the pairing symmetry. So far, we only consider the simple intraorbital pairing. In recent theoretical investigations, interorbital and interband pairing in iron pnictides are widely studied and deliver many interesting and exotic phenomena\cite{Dagotto09,Chubukov09}. In principle, the methods we developed here can be applied to the interorbital or interband pairing but the detail analysis remains open and requires further investigations.

We acknowledge supports from the National Science Council in Taiwan through grant NSC-97-2112-M-007-022-MY3 and partial financial supports from the National Center for Theoretical Sciences in Taiwan.

\end{document}